\documentclass[11pt,reqno]{JHEP3}
\usepackage{amsmath,amssymb,longtable}
\title{The RR charges of the A-type Gepner models}  
\author{Claudio Caviezel, Stefan Fredenhagen and 
Matthias R.\  Gaberdiel\\  
Institut f{\"u}r Theoretische Physik, ETH-H{\"o}nggerberg\\
CH--8093 Z{\"u}rich, Switzerland\\ E-mail:
\email{cclaudio@student.ethz.ch}, 
\email{stefan@itp.phys.ethz.ch}, \email{gaberdiel@itp.phys.ethz.ch}}
\abstract{It is shown that the RR charges of Gepner models are not all
accounted for by the usual tensor product and permutation
branes. In order to characterise the missing D-branes we study the
matrix factorisation approach to the description of D-branes
for Gepner models. For each of the A-type models we identify a set of 
matrix factorisations whose charges generate the full lattice of
quantised charges. The additional factorisations that are required 
correspond to generalised permutation branes.} 
\preprint{}
\begin{document}

\section{Introduction}

Most phenomenologically relevant string compactifications involve
Calabi-Yau manifolds. From a conformal field theory point of view,
string compactifications on a Calabi-Yau manifold can be described (at
least at some specific points in their moduli space) in terms of
Gepner models
\cite{Gepner:1987qi,Gepner:1987vz,Greene:1988ut,Witten:1993yc}. Gepner  
models are certain orbifolds of tensor products of minimal N=2
superconformal models. Since in most phenomenological string
constructions D-branes play a crucial r{\^o}le, it is  an important
problem to construct and understand the D-branes for these models.   

A certain class of D-branes for Gepner models can be relatively easily
constructed: these are the so-called tensor product or
Recknagel-Schomerus (RS) D-branes \cite{Recknagel:1998sb} that
preserve the different N=2 superconformal algebras separately. A
slight generalisation of this construction involves D-branes that
preserve the different N=2 superconformal algebras up to a
permutation, the so-called permutation branes
\cite{Recknagel:2002qq}. It is then an interesting (and obvious)
question to ask whether these different constructions account already
for all the RR charges of these models. As we shall see, this is in
general not the case. For example, of the $147$ A-type
Gepner models corresponding to six-dimensional Calabi-Yau manifolds
there are $31$ where the tensor product and permutation branes do not
couple to all RR ground states.  This analysis can be performed
directly in conformal field theory.  
\smallskip

Recently, a beautiful and simple characterisation of the 
(topological aspects of) B-type D-branes in N=2 theories has been 
proposed in unpublished work by Kontsevich. According to this idea
one can characterise these D-branes in terms of matrix factorisations
of the superpotential. The Kontsevich proposal has been supported in
a number of papers 
\cite{Kapustin:2002bi,Brunner:2003dc,Kapustin:2003ga,Lazaroiu:2003zi,
Herbst:2004ax} by analysing the possible B-type boundary terms for the
Landau-Ginzburg action. For the case of the N=2 minimal models the 
Landau-Ginzburg results have also been shown to agree with the results
obtained using conformal field theory methods 
\cite{Brunner:2003dc,Kapustin:2003rc,Hori:2004bx,
Brunner:2005fv,Brunner:2005pq,Enger:2005jk}.
Furthermore, these techniques have been used to study D-branes on
Calabi-Yau manifolds 
\cite{Ashok:2004zb,Ashok:2004xq,Hori:2004ja,Brunner:2005fv,Enger:2005jk}. 

In this paper we want to use the matrix factorisation approach to
understand which D-branes actually generate the RR charges for a
certain class of Gepner models, the A-type models that
correspond to the Fermat type Calabi-Yau manifolds. Unlike in
conformal field theory where (so far) only tensor product and permutation
branes are accessible, the matrix factorisations of a wide class of 
D-branes can be easily constructed and analysed. This approach
therefore allows us to characterise the matrix factorisations
that generate all the charges for these Gepner models.\footnote{In
this paper we shall only consider B-type D-branes that couple to the
even cohomology charges. Using mirror symmetry, a similar analysis
should be possible for the A-type D-branes.} We shall find
that in addition to the factorisations that correspond to tensor
product and permutation branes, only one new class of factorisations
is required in order to account for the full quantised lattice of RR
charges. By analogy with the identification of \cite{Brunner:2005fv},  
we can identify the new factorisations with `generalised permutation
branes' that seem to exist whenever the (shifted) levels of the N=2
minimal models contain a non-trivial common factor. These D-branes are
therefore very reminiscent of the generalised permutation branes 
of products of WZW models for which evidence was recently found in
\cite{Fredenhagen:2005an}. 
\bigskip

The paper is organised as follows. In section~2 we describe the 
RR states that carry the even cohomology charges for all A-type 
Gepner models. We then identify explicitly the Gepner models whose RR 
charges cannot be accounted for in terms of the usual (B-type)
tensor product or permutation D-branes. In section~3, we briefly
review the matrix factorisation approach to the analysis of B-type 
D-branes for Gepner models, and analyse a new class of rank $1$
factorisations that appear whenever two exponents in
the superpotential $W=\sum_i x_i^{h_i}$
contain a non-trivial common factor. We also determine the associated
intersection matrices. In section~4 we use these results (as well as
known formulae for the intersection matrices of tensor product and
permutation factorisations) to identify, for all $147$     
A-type Gepner models, which matrix factorisations are required to
account for all the RR charges. Our results are briefly summarised in
section~5, and the complete list of the required matrix factorisations
is given in the appendix.

\section{Ramond-Ramond ground states in Gepner models}

Let us begin by setting up some notation for the description of the RR
ground states in Gepner models \cite{Gepner:1987qi}. A more
comprehensive introduction to Gepner models can be found in
\cite{Gepner:1989gr}; our conventions are explained in more detail for
example in \cite{Recknagel:1998sb,FSW,Brunner:2005fv}. 

\subsection{Minimal models}
Gepner models are orbifolds of tensor products of N=2 supersymmetric
minimal models. The central charge of an N=2 minimal model is  
\begin{equation}
c=\frac{3\, k}{k+2} \ ,
\end{equation}
where $k$ is a positive integer. The representations of the bosonic
subalgebra of the N=2 superconformal algebra are labelled by triples
$(l,m,s)$ of integers, where $l$ takes the values $l=0,\dotsc ,k$
and $m$ and $s$ are defined modulo $2k+4$ and $4$,
respectively. The three integers have to obey $l+m+s=0\mod 2$. 
Furthermore there is an identification 
\begin{equation}
(l,m,s)\sim (k-l,m+k+2,s+2) \ .
\end{equation}
The conformal weight $h$ and the U(1)-charge $q$ of states in the
representation $(l,m,s)$ are given by~
\begin{equation}\label{weightcharge}
h (l,m,s)= \frac{l (l+2)-m^{2}}{2k+4} +\frac{s^{2}}{8} \mod \mathbb{Z}
, \qquad q (l,m,s)= \frac{s}{2}-\frac{m}{k+2}  \mod
2\mathbb{Z} .
\end{equation}
Representations with $s$ even belong to the
Neveu-Schwarz sector, while those with $s$ odd belong to the Ramond
sector. 

Ramond ground states are characterised by the property that their
conformal weight $h$ obeys 
\begin{equation}
h=\frac{c}{24} \ .
\end{equation}
One can easily show that a sector $\mathcal{H}_{(l,m,s)}$ contains a Ramond
ground state precisely if it is of the form $\mathcal{H}_{(l,l+1,1)}$
or $\mathcal{H}_{(l,-l-1,-1)}$. In this case, the 
formulae~\eqref{weightcharge} are valid exactly (not only up to
integers). Furthermore, we note that $(l,l+1,1)\sim (l',-l'-1,-1)$
with $l'=k-l$.

\subsection{Gepner models}
In this paper we are interested in A-type Gepner models 
that describe six-dimensional Calabi-Yau compactifications. These
are constructed starting with (at most) five N=2 minimal models with
A-type modular invariants whose central charges add up to
$c=9$.\footnote{There exist another $21$ Gepner models with $c=9$ that
involve more than five factors. As is explained in
\cite{Greene:1988ut}, these describe products of tori and K3s. We
shall not consider them in this paper.}  From a geometrical point of
view these Gepner models correspond to Fermat type Calabi-Yau
manifolds in $\mathbb{CP}^4$. There are precisely $147$ such models
that have been classified a long time ago   
\cite{Lynker:1988qi,Candelas:1989hd,Klemm:1992bx,Kreuzer:1992da}.

In the Gepner model construction the tensor product of the minimal
models is then projected onto states whose total U(1) charge is
integral. The corresponding orbifold is in fact a simple current
orbifold where the simple current $J$ acts on the representation
$(l_{i},m_{i},s_{i})$ of the $i^{\text{th}}$ minimal model by   
\begin{equation}
J: (l_{i},m_{i},s_{i})\longrightarrow (l_{i},m_{i}+2,s_{i}) \ .
\end{equation}
This is to say that in the $n^{\text{th}}$ twisted sector the right-movers
are shifted by $J^n$ relative to the left-movers. The order $H$ of
the orbifold group is given by the least common multiple of the
shifted levels, $H=\text{lcm} \{k_{i}+2 \}$.   

\noindent The Ramond-Ramond sectors that appear in the orbifold theory
are thus of the form
\begin{equation}\label{RRsectors}
\bigotimes_{i=1}^{5} 
\mathcal{H}_{(l_{i},m_{i}+n,s_{i})}\otimes
\bar{\mathcal{H}}_{(l_{i},m_{i}-n,\bar{s}_{i})} \ ,
\end{equation}
where $n=0,1,\dotsc ,H-1$ denotes the twisted sector, and
$s_{i},\bar{s}_{i}$ take values $-1,1$.  Because of the orbifold
projection the labels $m_{i}$ are subject to the integrality condition  
\begin{equation}\label{intcond}
\sum_{i=1}^{5} \frac{m_{i}}{k_{i}+2} \in \mathbb{Z}\ .
\end{equation}
[Recall that $\sum_{i=1}^5 \frac{1}{k_i+2} = 1$.] The labels $s_{i}$
and $\bar{s}_{i}$ in~\eqref{RRsectors} are restricted in a way that
depends on the GSO-projection. The GSO-projection that is
compatible with B-type RR states is 
\begin{equation}\label{GSO}
\sum_{i=1}^{5} \Big( \frac{s_{i}}{2}+\frac{\bar{s}_{i}}{2}\Big) \in 
2\mathbb{Z} \ .
\end{equation}
In fact, it is easy to see that all RR sectors~\eqref{RRsectors} that
satisfy the integrality condition~(\ref{intcond}) as well as the
B-type condition $q_L+q_R=0$ on the left-/right-moving U(1)-charge
necessarily obey~(\ref{GSO}).  

The RR ground states of a Gepner model are then those states that are
RR ground states in each minimal model factor.

\subsection{RR-charges of D-branes in Gepner models}
The RR charges we are interested in correspond to the even-dimensional
cohomology of the Calabi-Yau manifold, whose charges are carried by
the B-type D-branes. These D-branes can only couple to sectors whose
total left-moving U(1)-charge $q_{L}$ is opposite to its
right-moving charge, $q_{L}=-q_{R}$. A special class of such branes
are the tensor product branes \cite{Recknagel:1998sb} which 
preserve the individual N=2 superconformal symmetries of the five
minimal models. They can therefore only couple to sectors with
opposite U(1)-charges in each factor,
$q^{(i)}_{L}=-q^{(i)}_{R}$. Another class of branes whose 
conformal field theory description is known are the permutation
branes \cite{Recknagel:2002qq} which can occur if two or more levels
are equal. In the simplest case of two equal levels $k_{1}$ and
$k_{2}$, the permutation brane can only couple to sectors satisfying
$q^{(1)}_{L}=-q^{(2)}_{R}$ and $q^{(2)}_{L}=-q^{(1)}_{R}$. If there
are three or more equal levels, more permutations than just single
transpositions are possible, but as will become apparent below, these
will not be relevant for our considerations. 
\medskip

In this section we want to analyse whether these two types of branes
already account for all the RR charges of the above Gepner models. As 
is clear from the above discussion, a tensor product brane can only
couple to the RR ground states in the sectors 
\begin{equation}\label{tensorground}
\bigotimes
\mathcal{H}_{(l_{i},l_{i}+1,1)}\otimes
\bar{\mathcal{H}}_{(l_{i},-l_{i}-1,-1)}\ ,
\end{equation}
where $m_{i}=l_{i}+1$ have to obey the integrality
condition~\eqref{intcond}.  Generically, the states in 
(\ref{tensorground}) will come from the $n^{\text{th}}$ twisted sector
where $n$ satisfies $n=l_{i}+1\mod k_{i}+2$. In the special situation
when $k_{i}$ is even and $l_{i}=\frac{k_{i}}{2}$, also $n=0 \mod k_{i}+2$ is
allowed. For some Gepner models, all RR ground states are of this
form, and thus all RR charges can be generated by tensor product 
branes. 

There are however a number of examples for which some of the RR
charges can only be accounted for in terms of permutation branes.
(The simplest example is the theory with $(k=7)^3 (k=1)^2$, as was
already mentioned in \cite{Brunner:2005fv}.) A permutation brane
in two factors (for ease of notation we are ignoring here the
remaining three factors) couples to RR ground states in the sectors
\begin{equation}\label{permground}
\big( \mathcal{H}_{(l,l+1,1)} \otimes
\bar{\mathcal{H}}_{(l,- l-1, - 1)}\big) \otimes 
\big( \mathcal{H}_{(l, l+1, 1)} \otimes 
\bar{\mathcal{H}}_{(l, - l-1,- 1)}\big) \ .
\end{equation}
It is easy to see that a permutation brane involving more than two 
factors does not couple to any new RR ground states; such branes only
couple to RR ground states that can be built from the ground states in
\eqref{permground} and~\eqref{tensorground}.
\smallskip

It is natural to ask whether there are RR ground states to which 
neither tensor product nor permutation branes can couple. As we
have mentioned above, the B-type condition requires that the total
U(1) charges satisfy $q_{L}=-q_{R}$. A RR ground state 
in a sector~\eqref{RRsectors} has to have the same value for $l_i$ in
the left-moving and right-moving part, so it satisfies 
$q^{(i)}_{L}=\pm q^{(i)}_{R}$. In a given sector we now order the
factors such that the first $r$ factors satisfy
$q^{(i)}_{L}=q^{(i)}_{R}$ and the last $(5-r)$ factors satisfy
$q^{(i)}_{L}=-q^{(i)}_{R}$. There is an ambiguity whenever some of the
$q^{(i)}_{L}$ are zero. This allows us to always choose $r$ to be
even. Namely, assume that $r$ is odd and $q^{(i)}_{L}\not= 0$ for all
$i$. Then necessarily $s_{i}=\bar{s}_{i}$ for $i=1,\dotsc ,r$ and
$s_{j}=-\bar{s}_{j}$ for $j=r+1,\dotsc ,5$. This contradicts the
GSO condition~\eqref{GSO}. So if $r$ is odd there is at least one
factor with $q^{(i)}_{L}=0$ and we can lower or raise $r$ to an even
value. 

The tensor product case corresponds to the case $r=0$, so we are only
interested in $r=2,4$. The case $r=4$ actually never occurs which
can be seen as follows: From $q_L=-q_R$ and $q_L^{(5)}=-q_R^{(5)}$ it follows 
immediately that 
\begin{equation}
q_{L}-q^{(5)}_{L}= - (q_{R}-q^{(5)}_{R}) \ . 
\end{equation}
Since $q_L^{(i)}=q_R^{(i)}$ for $i=1,\ldots,4$ we also 
have $q_{L}-q^{(5)}_{L}= + (q_{R}-q^{(5)}_{R})$, and thus 
$q_{L}=q^{(5)}_{L}$. The integrality condition~\eqref{intcond} implies
that $q_{L}$ is a half-integer. On the other hand $q^{(5)}_{L}$
is of the form
\begin{equation}
q^{(5)}_{L}=\frac{1}{2}-\frac{l_{5}+1}{k_{5}+2}
\end{equation}
which is never half-integer. The case $r=4$ can thus be
excluded. 

This leaves us with the case $r=2$. The charges of the individual
factors then satisfy
\begin{align}
q^{(1)}_{L}&=q^{(1)}_{R} & q^{(2)}_{L}&=q^{(2)}_{R} &
q^{(i)}_{L}&=-q^{(i)}_{R} \ , \quad i= 3,4,5\  .
\end{align}
Together with the B-type condition $q_{L}=-q_{R}$ this implies that 
\begin{equation}\label{permcondI}
q^{(1)}_{L}+q^{(2)}_{R}=q^{(2)}_{L}+q^{(1)}_{R}=0 \ .
\end{equation}
Without loss of generality we may assume that the first left-moving
Ramond ground state is in the sector
$\mathcal{H}_{(l_{1},l_{1}+1,1)}$, while the second is in the sector
$\mathcal{H}_{(l_{2},-l_{2}-1,-1)}$. To achieve~\eqref{permcondI} 
we thus need 
\begin{equation}\label{permcondII}
\frac{l_{1}+1}{k_{1}+2}=\frac{l_{2}+1}{k_{2}+2}\ .
\end{equation}
This condition can only be satisfied if the shifted levels $k_{1}+2$
and $k_{2}+2$ have a non-trivial common divisor $d$,
\begin{equation}
d=\gcd (k_{1}+2,k_{2}+2) \ .
\end{equation}
For $d=2$, however, the only solution to~\eqref{permcondII} is
$l_{i}=k_{i}/2$ which implies $q^{(i)}_{L/R}=0$. Then we are back to
the tensor product case, so we can assume $d>2$. If the levels are
equal, $k_{1}=k_{2}$, we recover the RR ground states that couple to
permutation branes.  New RR ground states can only be expected if
$d>2$ and $k_{1}\not= k_{2}$. These RR ground states can only exist in
sectors whose twist $n$ is trivial in the first two factors, so
\[
n=0 \mod r_{1}r_{2}d \ , 
\]
where $k_{i}+2=r_{i}d$.
The procedure to find all of these states is now as follows. In each 
Gepner model we look for levels which are different and have a
non-trivial common divisor $d>2$. Then we investigate the twisted
sectors with a twist $n\in r_{1}r_{2}d\cdot \mathbb{Z}$ and see whether there
exists a RR ground state satisfying~\eqref{permcondII} and the integrality
condition~\eqref{intcond}. 

The analysis can be implemented on a computer. Out of the $147$ Gepner
models of A-type, there are $31$ models with RR ground states of type
$r=2$ which do not couple to tensor product or permutation branes ---
these have been collected in table~\ref{tab:RRgs}. In the simplest
example the levels are $(6,6,4,2,1)$. The relevant two levels are
$k_{3}=4$ and $k_{5}=1$ whose shifted levels have a common divisor
$d=3$. There are RR ground states of the type $r=2$ in the
$6^{\text{th}}$ and $18^{\text{th}}$ twisted sectors. For example, for
$n=6$ the relevant ground state appears in the sector
\begin{equation}
\begin{array}{lccccccccl}
\hbox{L} \qquad  & (1,-2,-1)_6 & \otimes  & (1,-2,-1)_6 
& \otimes & (1,2,1)_4 & \otimes & (1,2,1)_2 
& \otimes & (0,-1,-1)_1 \nonumber \\
\hbox{R}  \qquad & (1,2,1)_6 & \otimes & (1,2,1)_6 & 
\otimes & (1,2,1)_4 & \otimes & (1,-2,-1)_2 
& \otimes & (0,-1,-1)_1 \ , \nonumber 
\end{array}
\end{equation}
where the first line describes the left- and the second line the 
right-moving representations, and the indices denote the levels
$k_i$. One easily checks that this combination of representations
appears in the $n=6$ sector ({\it i.e.} is of the form
(\ref{RRsectors}) with $n=6$), and  that it satisfies the integrality
condition (\ref{intcond}). On the other hand, it is clear that this RR
ground state cannot couple to any tensor product or permutation brane.  

\small
\renewcommand{\arraystretch}{1.1}
\begin{longtable}{lcl}
\caption{\label{tab:RRgs}\normalsize The Gepner models with RR ground states of
type $r=2$ in the factors $i,j$. $n$ denotes the sectors in which they
appear.}\\
\hline
Levels & ($i,j$) & $n$\\
\hline\hline
\endfirsthead
\caption{\normalsize (continued)}\\
\hline
Levels & ($i,j$) & $n$\\
\hline\hline
\endhead
(6, 6, 4, 2, 1)& (3, 5)& (6, 18)\\
(8, 4, 3, 3, 1)& (2, 5)& (6, 12, 18, 24)\\
(13, 8, 2, 2, 1)& (1, 5)& (15, 45)\\
(16, 7, 2, 2, 1)& (2, 5)& (9, 27)\\
(18, 4, 3, 2, 1)& (2, 5)& (6, 18, 42, 54)\\
(18, 10, 3, 1, 1)& (1, 3)& (20, 40)\\
(18, 18, 13, 1, 0)& (3, 4)& (15, 45)\\
(26, 19, 2, 1, 1)& (1, 3)& (28, 56)\\
(26, 19, 10, 1, 0)& (2, 4)& (21, 63)\\
(28, 8, 3, 1, 1)& (2, 3)& (10, 20)\\
(28, 18, 2, 1, 1)& (2, 3)& (20, 40)\\
(33, 13, 12, 1, 0)& (2, 4)& (15, 45, 75, 135, 165, 195)\\
(34, 34, 7, 1, 0)& (3, 4)& (9, 27)\\
(43, 28, 7, 1, 0)& (3, 4)& (9, 27, 63, 81)\\
(43, 34, 3, 2, 0)& (1, 3)& (45, 135)\\
(46, 14, 2, 1, 1)& (2, 3)& (16, 32)\\
(52, 25, 7, 1, 0)& (3, 4)& (9, 45)\\
(58, 13, 10, 1, 0)& (2, 4)& (15, 45)\\
(70, 22, 7, 1, 0)& (3, 4)& (9, 27, 45, 63)\\
(86, 31, 6, 1, 0)& (2, 4)& (33, 99, 165, 231)\\
(89, 76, 5, 1, 0)& (1, 3)& (91, 455)\\
(97, 20, 7, 1, 0)& (3, 4)& (9, 27, 45, 63, 81, 117, 135, 153, 171, 189)\\
(98, 23, 3, 2, 0)& (2, 3)& (25, 75)\\
(108, 13, 9, 1, 0)& (2, 4)& (15, 45, 75, 105, 135, 195, 225, 255, 285, 
   315)\\
(124, 19, 7, 1, 0)& (2, 4)& (21, 105)\\
                  & (3, 4)& (9, 27, 45, 81, 99, 117)\\
(178, 18, 7, 1, 0)& (3, 4)& (9, 27, 45, 63, 81, 99, 117, 135, 153, 171)\\
(214, 25, 6, 1, 0)& (2, 4)& (27, 81, 135, 189)\\
(236, 49, 5, 1, 0)& (2, 4)& (51, 153, 255, 459, 561, 663)\\
(292, 47, 5, 1, 0)& (2, 3)& (49, 245)\\
(340, 17, 7, 1, 0)& (3, 4)& (9, 27, 45, 63, 81, 99, 117, 135, 153, 189,\\
                  &       & \ 207, 225, 243, 261, 279, 297, 315, 333)\\
(628, 43, 5, 1, 0)& (2, 4)& (45, 135, 225, 405, 495, 585)\\[0.1cm]
\hline\hline
\end{longtable}
\normalsize 

Our analysis thus shows that in general the tensor product or permutation
branes do not account for all the charges. Furthermore, it suggests
that the only additional D-brane construction that is required to
account for these charges involves two factors whose shifted levels
have a non-trivial common factor. This is very reminiscent of the
generalised permutation branes for factors of SU(2) WZW models for
which evidence was recently found in \cite{Fredenhagen:2005an}.

In the following sections we shall analyse the same problem using the 
matrix factorisation point of view. We shall be able to reproduce the
above results, but we shall also be able to identify the matrix
factorisations that will actually account for all the RR charges. In 
particular, we shall find that the new factorisations that are
required are indeed a natural generalisation of the factorisations 
that correspond to permutation branes.

\section{Matrix factorisations}

D-branes in Gepner models can also be analysed in terms of  
orbifolds of Landau-Ginzburg models that flow in the IR to the
relevant superconformal field theory. In particular, B-type D-branes
in Landau-Ginzburg models can be described in terms of matrix 
factorisations. This approach goes back to unpublished work of
Kontsevich, and the physics interpretation of it was given in  
\cite{Kapustin:2002bi,Brunner:2003dc,Kapustin:2003ga,Kapustin:2003rc, 
Lazaroiu:2003zi,Herbst:2004ax}; for a good review of this material see
for example~\cite{Hori:2004zd}.   

\subsection{Generalities}

According to Kontsevich's proposal, D-branes in Landau-Ginzburg models
correspond to matrix factorisations of the superpotential $W$,
\begin{equation}\label{factorize}
E\, J = J \, E = W\cdot {\bf 1}\ ,
\end{equation}
where $E$ and $J$ are $r\times r$ matrices. This condition can be more
succinctly written as 
\begin{equation} \label{Qdef}
Q^2=W\cdot \mathbf{1}\ , \qquad \mbox{where} \qquad 
Q=\left(\begin{array}{cc} 0 & J \\ E & 0 \end{array}\right) \ . 
\end{equation}
The theories that are of interest to us\footnote{The matrix
factorisation description of supersymmetric D-branes also applies to
more general classes of Landau-Ginzburg models --- see for example 
\cite{Mattik:2005du} for a recent discussion in the context of the
N=2 sine-Gordon model.} have a quasi-homogeneous superpotential
$W(x_i)$,
\begin{equation}
W\left(\lambda^{w_1}x_1,\ldots,\lambda^{w_n}x_n\right)=
\lambda^{H} \, 
W\left(x_1,\ldots,x_n\right)\quad \mbox{for}\quad 
\lambda \in \mathbb{C}  \label{eq30} \ ,
\end{equation}
ensuring that the bulk theory is superconformal rather
than just supersymmetric. More specifically, the A-type Gepner models 
correspond to polynomials of Fermat type; this means that 
all $w_i$ divide $H$ so that $W$ takes the form (we restrict our
discussion to the case of five variables)
\begin{equation}\label{super}
W=x_1^{h_1}+x_2^{h_2}+x_3^{h_3}+x_4^{h_4}+x_5^{h_5}\ ,
\end{equation}
where $h_i=H/w_i$. 

The factorisations $Q$ that are in one-to-one correspondence
to the {\it superconformal} D-branes must also respect this U(1)
symmetry; this implies that the entries of $Q$ must be polynomials in
the $x_i$, and  furthermore, that there exist matrices $R(\lambda)$ so
that   
\begin{equation}
R(\lambda) \, Q(\lambda^{w_i} x_i) \, R(\lambda)^{-1} 
= \lambda^{\frac{H}{2}} \,  Q(x_i)  \ .
\end{equation}
The matrices $R(\lambda)$ as well as $R(\lambda)^{-1}$ may depend on
the $x_i$ in a polynomial way.
\smallskip

To make contact with the Gepner models we have to consider an orbifold
of the Landau-Ginzburg model, where the orbifold group $\mathbb{Z}_H$
acts on the variables $x_i$ as 
\begin{equation}
x_i \mapsto \omega^{w_i} x_i\ , \qquad
\hbox{where} \qquad
\omega = e^{\frac{2\pi i }{H}} \ .
\end{equation}
This orbifold action must also be implemented on the open
strings, and thus we need to choose an action of the orbifold group on
$Q$. More precisely, we need to choose a matrix $\gamma$ (that
together with its inverse is polynomial\footnote{Note that in all
examples which we shall discuss $\gamma$ is constant.} in the $x_i$) 
such that $Q$ satisfies the equivariance condition 
\begin{equation}
\gamma\, \,
Q\left(\omega^{w_i} x_i\right)\, \gamma^{-1}
=Q(x_i)\ , \label{eq31}  
\end{equation}
where $\gamma^H = {\bf 1}$. If such a $\gamma$ exists, it is not
unique, since we can always multiply a given $\gamma$ by an
$H^{\text{th}}$ root of unity. A D-brane in the
orbifold theory is thus characterised by $Q$, together with a choice
of representation $\gamma$. 
\smallskip

Given two D-branes described by $(Q, \gamma )$ and 
$(\widehat{Q}, \hat{\gamma})$, the open string
spectrum between them is determined by a suitably defined equivariant
cohomology. More precisely, the bosons are the maps of the form 
\begin{equation}
\phi=\left(\begin{array}{cc} \phi_0 & 0 \\ 0 & \phi_1
\end{array}\right) 
\end{equation}
that are invariant under the orbifold action, {\it i.e.} 
\begin{equation}
\phi(x_i) = \hat{\gamma} \, \, 
\phi\left(\omega^{w_i} x_i\right) \,\,  \gamma ^{-1} \ ,
\end{equation}
and satisfy the BRST-closure condition
\begin{equation}
\widehat{Q}\, \phi = \phi \, Q \ .
\end{equation}
These bosons are only considered modulo the BRST-exact solutions
\begin{equation}
\tilde{\phi} = \widehat{Q} \, t + t \, Q \ ,  \qquad 
\hbox{where} \qquad 
t = \left(\begin{array}{cc} 0 & t_1 \\ t_0 & 0 \end{array}\right) 
\end{equation} 
describes a fermion. Similarly, the fermions are the invariant maps
$t$ that satisfy the BRST-closure condition 
$\widehat{Q} \, t + t\, Q  = 0$, modulo the BRST-exact solutions,  
$\tilde{t} = \widehat{Q} \phi - \phi \, Q$. Finally, the index between
two such D-branes is the number of bosonic states minus the number of
fermionic states. For example, for a single minimal model, the
self-intersection matrix for the matrix factorisation 
for which $J$ is linear in $x$ is simply \cite{Ashok:2004zb}
\begin{equation}
I = ({\bf 1} - G^{-w})\ ,
\end{equation}
where $G$ is the $H$-dimensional shift matrix, and $w=H/h$. This
$H$-dimensional intersection matrix accounts for the $H$ different
choices for the matrix $\gamma$ in (\ref{eq31}) that are obtained from
a given $\gamma$ by multiplication by an $H^{\text{th}}$ root of unity.
\smallskip

Finally, as was explained in \cite{Ashok:2004zb,Hori:2004ja}, one can
tensor matrix factorisations together: if $Q_1$ and $Q_2$ are matrix
factorisations for the superpotentials $W_1$ and $W_2$, respectively,
the tensor product factorisation $Q_1\widehat{\otimes} Q_2$ is a matrix 
factorisation for $W_1+W_2$. Furthermore, the intersection matrix for
the tensor product factorisation is just the product of the separate
intersection matrices. For example, for the five-fold tensor product
factorisation of the superpotential (\ref{super}), the
self-intersection matrix of the factorisations for which each $J_i$ is
linear is simply  
\begin{equation}\label{RSind}
I_{{\rm RS}} = \prod_{i=1}^{5} ({\bf 1} - G^{-w_i})\ .
\end{equation}
This then agrees with the Witten index of the open string spectrum of
the tensor product (or Recknagel-Schomerus) branes with $L_i=0$ 
\cite{Ashok:2004zb}. 

\subsection{Generalised permutation factorisations}

It was shown in \cite{Brunner:2005fv} that the transposition branes
({\it i.e.} the branes where the permutation is just a transposition
between the two factors $k_1=k_2$ say) correspond to tensor products
of factorisations that involve the rank $1$ factorisation in the first
two factors ($d=k_1+2=k_2+2$) 
\begin{equation}
W_{12} = x_1^d + x_2^d = J\, E \ , \qquad
\hbox{where} \qquad
J = \prod_{\eta \in {\cal I}} (x_1 - \eta x_2)\ , \qquad
E = \prod_{\eta' \not\in {\cal I}} (x_1 - \eta' x_2) \ ,
\end{equation}
as well as the usual factorisations for $x_3,x_4$ and $x_5$. Here the
$\eta$'s run over the $d$ different $d^{\text{th}}$ roots of $-1$, and
${\cal I}$ denotes some (suitable) subset of these roots. Such
factorisations were first considered in~\cite{Ashok:2004zb}. The
generalisation for higher order permutations was found in
\cite{Enger:2005jk}.

It is relatively obvious how this construction can be generalised to
the case where the two exponents are not the same, but only contain a
non-trivial common factor. To set up notation, let us consider a
superpotential in two factors
$W=x_1^{h_1}+x_2^{h_2}=x_1^{dr_{1}}+x_2^{dr_{2}}$,
where $d\geq 2$ is the greatest common divisor of $h_1$ and $h_2$,
$d=\gcd (h_1,h_2)$. We can obviously factorise
\begin{equation}
W=x_1^{dr_{1}}+x_2^{dr_{2}}=
\prod_{\eta}(x_1^{r_{1}}-\eta x_2^{r_{2}})\ , \label{eq22}
\end{equation}
where $\eta$ is in turn each of the $d$ different $d^{\text{th}}$ roots of
$-1$. Then we can define a rank $1$ factorisation by taking $J$ to be
the product of some of these factors, with $E$ being the product of the
remaining factors. We shall call these rank $1$ factorisations
{\it generalised permutation factorisations}, and we shall sometimes
denote them by $(\widetilde{12})$. By the same arguments as
in \cite{Brunner:2005fv} one can show that the factorisations where
$J$ or $E$ contains more than one factor can be obtained as bound
states of those where either $J$ or $E$ is a single factor. For the
analysis of the charges it should therefore be sufficient to consider
these factorisations only, and this is indeed what we shall find. 
In the following we thus concentrate on factorisations for which
$J$ consists of a single factor, $J=(x_1^{r_1}- \eta
x_2^{r_2})$. They will be denoted by $Q_{\eta}$. 

For the analysis of the charges it is important to determine the
corresponding open string spectra, and in particular the index. This
will be done next.

\subsection{Spectra and indices \label{spectra}}

The calculations to determine the spectra and indices are all
relatively straightforward, so we shall only explain them in one
example and give the results for the remaining cases. The simplest
case is the open string spectrum between two branes corresponding to
generalised permutation factorisations.

\subsubsection{Spectrum of generalised permutation factorisations}
The discussion of the spectrum between two factorisations $Q_{\eta}$
and $\widehat{Q}_{\hat{\eta}}$ depends on whether $\eta$ and
$\hat{\eta }$ coincide or not, so we shall distinguish these two
cases.
\subsubsection*{The case $\eta \not= \hat{\eta}$}
The BRST-closure condition for the bosons 
$\widehat{Q}_{\hat{\eta}}\, \phi- \phi\,Q_{\eta} = 0$ gives
\begin{align}
\phi_1 \, (x_1^{r_{1}}-\hat{\eta} x_2^{r_{2}})
- \phi_0 \, (x_1^{r_{1}}-\eta x_2^{r_{2}}) & = 0\nonumber \\
\phi_0\, \prod\limits_{\hat{\eta}' \neq \hat{\eta}}
(x_1^{r_{1}}-\hat{\eta}' x_2^{r_{2}})
-\phi_1 \, 
\prod\limits_{\eta' \neq \eta}(x_1^{r_{1}}-\eta' x_2^{r_{2}})
& = 0\ .
\end{align}
Since $\eta \neq \hat{\eta}$, the unique solution is
\begin{equation}\label{bossol}
\phi_0 = a \left(x_1^{r_{1}}-\hat{\eta} x_2^{r_{2}}\right) \qquad
\mbox{and}\qquad 
\phi_1 = a \left(x_1^{r_{1}}-\eta x_2^{r_{2}}\right)\ , 
\end{equation}
where $a \in \mathbb{C}[x_1,x_2]$ is an arbitrary polynomial. The
bosonic operator is BRST-exact if 
$\phi=\widehat{Q}_{\hat{\eta}} \,t + t \, Q_{\eta}$ for an arbitrary 
fermionic operator $t$. It is easy to see that every solution
(\ref{bossol}) is BRST-trivial and thus there are no bosons 
propagating between two such branes.

\noindent For the fermions, the BRST-closure condition gives
\begin{align}
t_0\, \left(x_1^{r_{1}}-\hat{\eta} x_2^{r_{2}}\right)
+t_1 \,
\prod\limits_{\eta' \neq \eta}\left(x_1^{r_{1}}-\eta'  x_2^{r_{2}}\right)
 &= 0 \nonumber \\
t_1\, \prod\limits_{\hat{\eta}' \neq \hat{\eta}}
\left(x_1^{r_{1}}-\hat{\eta}' x_2^{r_{2}}\right)
+t_0\, \left(x_1^{r_{1}}-\eta x_2^{r_{2}}\right)  & = 0\ .
\end{align}
This is solved by
\begin{equation}\label{BRSTsol}
t_1 \in \mathbb{C}[x_1,x_2] 
\quad\mbox{and}\quad 
t_0= -t_1 \prod\limits_{\eta' \neq \eta,\hat{\eta}}
\left(x_1^{r_{1}}-\eta' x_2^{r_{2}}\right)\ ,
\end{equation}
so the BRST-closed operators are determined by $t_{1}$.  
The fermionic operator is BRST-exact if
$t = \widehat{Q}_{\hat{\eta}} \, \phi - \phi\, Q_{\eta}$ 
for an arbitrary bosonic operator~$\phi$. This is the case if $t_1$
lies in the ring 
\begin{equation}
t_1\in\  <\left(x_1^{r_{1}}-\hat{\eta} x_2^{r_{2}}\right),
\left(x_1^{r_{1}}-\eta x_2^{r_{2}}\right)>\ .
\end{equation}
Representatives of the BRST cohomology can thus be chosen as
\begin{equation}
t^{(i,j)} \left(x_1,x_2\right)=
\left(\begin{array}{cc} 0 & x_1^i x_2^j \cr -x_1^i x_2^j 
\prod\limits_{\eta' \neq \eta,\hat{\eta}}
\left(x_1^{r_{1}}-\eta' x_2^{r_{2}}\right) 
\quad & 0 \end{array}\right)\ , \label{eq18}
\end{equation}
where
\begin{equation}
i=0,\ldots,r_{1}-1 \quad\mbox{and}\quad j=0,\ldots,r_{2}-1\ . 
\label{eq19} 
\end{equation}
The number of fermions propagating between the branes is therefore
$r_1 r_2$.  The equivariance condition~\eqref{eq31}
for $Q_{\eta}$ is satisfied by choosing $\gamma$ to be one of the 
matrices
\begin{equation}
\gamma_\mu  =
\left(\begin{array}{cc} \omega^{-\frac{H/d+\mu }{2}} & 0 \cr 0 
& \omega^{\frac{H/d-\mu }{2}} \end{array}\right)
\ , \label{eq57}
\end{equation}
where $\mu $ is an integer defined mod $2H$ that has the property that
$H/d+\mu $ is even (similarly one defines $\hat{\gamma}_{\hat{\mu}}$
with an integer $\hat{\mu}$). One then easily calculates that the
above fermionic operators transform as
\begin{equation}
\hat\gamma_{\hat{\mu }} \,\,
t^{(i,j)}\!\left(\omega^{w_l} x_l\right) \,
\gamma_\mu ^{-1} 
= \omega^{-\frac{H}{d}+w_1 i+w_2 j+\frac{\mu -\hat{\mu }}{2}} \,
t^{(i,j)}\!\left(x_l\right)\ ,
\end{equation}
where $i=0,\ldots,r_{1}-1$ and $j=0,\ldots,r_{2}-1$. We can thus
express the index in terms of the $H$-dimensional shift matrix $G$ 
\begin{equation}\label{indexetanotetap}
{I}_{(\widetilde{12})}=
-G^{-\frac{H}{d}}\left(\textbf{1}+G^{w_1}
+\cdots+G^{\left(r_{1}-1\right)w_1}\right)
\left(\textbf{1}+G^{w_2}+\cdots+G^{\left(r_{2}-1\right)w_2}\right)\ .
\end{equation}

\subsubsection*{The case $\eta=\hat\eta$}
For the case $\eta = \hat{\eta}$ one easily shows that there are no
fermions. For the bosons, one can take the representatives of the BRST
cohomology to be given by 
\begin{equation}
\phi^{(i,j)}(x_1,x_2)  
=\left(\begin{array}{cc} x_1^i x_2^j & 0 \\ 
0 & x_1^i x_2^j \end{array}\right)\ , \label{eq20}
\end{equation}
with 
\begin{equation}
i=0,\ldots,r_{1} (d-1)-1  \qquad \hbox{and} \qquad 
j=0,\ldots,r_{2}-1\ . \label{eq21}
\end{equation}
Thus the number of bosons propagating in the self-overlap is equal to 
$r_{1} r_{2} \left(d-1\right)$. With respect to the above
choice of $\gamma_\mu $ the bosonic fields transform as 
\begin{equation}
\hat{\gamma}_{\hat{\mu }}\, \phi^{(i,j)}\left(\omega^{w_l}x_l\right) \,
\gamma_\mu ^{-1} = 
\omega^{w_1 i+w_2 j+\frac{\mu -\hat{\mu }}{2}} 
\phi^{(i,j)}\left(x_l\right)\ ,
\end{equation}
and therefore the intersection matrix reads
\begin{equation}\label{indexetaeqetap}
{I}_{(\widetilde{12})}=
\left(\textbf{1}+G^{w_1}+\cdots
+G^{\left(r_{1}\left(d-1\right)-1\right)w_1}\right)
\left(\textbf{1}+G^{w_2}+\cdots+G^{\left(r_{2}-1\right)w_2}\right)\ .
\end{equation}
\medskip

Before we move on let us take a look at how our formulae simplify when
we consider the actual permutation case, $w_1=w_2$. Then we have
$d=h_1=h_2$ and thus $r_{1}=r_{2}=1$. If $\eta\neq\hat{\eta}$, the
intersection matrix~\eqref{indexetanotetap} takes the simple form  
\begin{equation}
I_{\left(12\right)}=-G^{-\frac{H}{d}}=-G^{-w_1}\qquad 
(\eta\neq \hat\eta)\ .
\end{equation}
For $\eta = \hat{\eta}$, on the other hand, the
intersection matrix~\eqref{indexetaeqetap} reads
\begin{equation}
{I}_{\left(12\right)}
=\left(\textbf{1}+G^{w_1}+\cdots+G^{\left(d-2\right)w_1}\right)\qquad
(\eta=\hat\eta)\ .
\end{equation}
The formulae for these special cases were already given in
\cite{Ashok:2004zb}.  

\subsubsection{Relative intersection forms} 
In order to determine the charges it is also important to calculate
the intersection form between these generalised permutation
factorisations and the tensor product factorisations. Using similar
techniques one finds that there is always one boson and one fermion
between these two branes. Furthermore, the U(1) charges are such that
the intersection matrix is 
\begin{equation}
{I}_{{\rm RS}- (\widetilde{12})}= 
G^{-\lfloor\frac{w_1}{2}\rfloor -\lfloor\frac{w_2}{2}\rfloor} 
\left(G^{\lfloor-\frac{H}{2d}\rfloor}
-G^{\lfloor\frac{H}{2d} \rfloor}\right) \ ,
\end{equation}
where $\lfloor n \rfloor$ denotes the Gauss bracket, {\it i.e.} the
greatest integer less or equal to $n$. 

For the case of the actual permutation branes these formulae reproduce 
again those of \cite{Ashok:2004zb}. We also note that the 
morphisms and the intersection matrices are in fact independent
of the choice of $\eta$.
\bigskip

Finally, we need two further classes of relative intersection forms 
for our analysis of the Gepner models. If the superpotential is of the
form 
\begin{equation}
W=x_1^{h_1}+x_2^{h_2}+x_3^{h_3} \ , \qquad
d_i=\gcd (h_i,h_{i+1}) \geq 2 \ , 
\end{equation}
then the intersection matrix between the two generalised permutation
factorisations corresponding to $(\widetilde{12})$ and
$(\widetilde{23})$ is \cite{Caviezel}
\begin{align}
{I}_{(\widetilde{12})- (\widetilde{23})} 
 & =  G^{-\lfloor\frac{H}{2d_{1}}\rfloor + 
    \lfloor -\frac{H}{2d_{2}}\rfloor
+\lfloor-\frac{w_1}{2}\rfloor 
- \lfloor \frac{w_3}{2} \rfloor} \nonumber \\
   & \qquad \times 
\left(\mathbf{1}+G^{w_2}+\cdots
+G^{w_2\left(\text{min}\left(r,s\right)-1\right)}\right)
\left(\mathbf{1}-G^{w_2\text{max}\left(r,s\right)}\right)
\ ,
\end{align}
where $r=h_2/d_1$ and $s=h_2/d_2$. 

If the superpotential is of the form 
\begin{equation}
W=x_1^{h_1}+x_2^{h_2}+x_3^{h_3} + x_4^{h_4}\ , \qquad
d_i=\gcd (h_i,h_{i+1}) \geq 2 \ , 
\end{equation}
and we define $r_1=h_2/d_1$, $r_2=h_3/d_2$, then the intersection
matrix between the two generalised permutation factorisations
$(\widetilde{12})(\widetilde{34})$ and $(\widetilde{23})$ is
\cite{Caviezel}  
\begin{align}
{I}_{(\widetilde{12})(\widetilde{34})- (\widetilde{23})} 
& =  G^{-\lfloor\frac{H}{2d_{1}}\rfloor 
- \lfloor\frac{H}{2d_{3}}\rfloor
+ \lfloor-\frac{H}{2d_{2}}\rfloor
+\lfloor-\frac{w_1}{2}\rfloor 
+ \lfloor -\frac{w_4}{2} \rfloor} \ 
\left(\mathbf{1}+G^{w_2}+\cdots
+G^{w_2(r_1-1)}\right) \nonumber \\
&  \qquad \times 
\left(\mathbf{1}+G^{w_3}+\cdots
+G^{w_3(r_2-1)}\right) \
\left(\mathbf{1}-G^{\frac{H}{d_3}} \right)
\ , 
\end{align} 
where we spell out for simplicity only the case where $d_1\geq d_2
\geq d_3$, since the other cases will not be relevant for our analysis
of the Gepner models.  

\subsection{RR charges from path integrals}

There is an alternative way to determine the intersection matrix by
computing directly the RR-charges of the factorisations, {\it i.e.}
the topological correlators of one bulk-field in the presence of a
boundary. For Landau-Ginzburg models with the boundary condition
described by a matrix factorisation, these correlators can be
determined using path integral methods~\cite{Kapustin:2003ga}. A
generalisation to orbifolds has been proposed
in~\cite{Walcher:2004tx}. Before we spell out the concrete expressions
of \cite{Walcher:2004tx}, we need to introduce some notation. We label
the RR ground states $|n;\alpha \rangle$ in the $n^{\text{th}}$
twisted sector by a 
multi-index $\alpha= (\alpha_{1},\dotsc ,\alpha_{u_{n}})$. Here,
$u_{n}$ is the number of factors for which $n=0 \mod h_{i}$ and we
assume that we have re-ordered the factors such that these are the
first $u_{n}$ ones. The state
$|n;\alpha \rangle$ is the ground state in the sector 
\begin{equation}
\bigotimes_{i=1}^{u_{n}} \mathcal{H}_{(\alpha_{i},-\alpha_{i}-1,-1)}
\otimes \bar{\mathcal{H}}_{(\alpha_{i},-\alpha_{i}-1,-1)} \otimes
\bigotimes_{j=u_{n}+1}^{5} \mathcal{H}_{(n_{j}-1,n_{j},1)} \otimes
\bar{\mathcal{H}}_{(n_{j}-1,-n_{j},-1)}\bigg|_{n_{j}=n\mod h_{i}} \ . 
\end{equation}
It is obtained from the state $|n;0\rangle$ by acting
with the field 
$\phi_{n}^{\alpha}=\prod_{i=1}^{u_{n}} x_{i}^{\alpha_{i}}$.

The charge of the factorisation $Q$ (together with a representation
$\gamma$ of the orbifold group) under the field corresponding to
$|n;\alpha\rangle$ is then given by~\cite{Walcher:2004tx} 
\begin{equation}\label{charge}
\text{ch}(Q,\gamma)(|n;\alpha\rangle)=\frac{1}{u_{n}!} 
\text{Res}_{W_n}(\phi_{n}^{\alpha} \,
\text{Str}[\gamma^{n}(\partial Q)^{\wedge u_{n}}]) \ .
\end{equation}
Here Str denotes the supertrace, {\it i.e.} the difference between the
trace of the upper left and the lower right $r\times r$ block of the  
$2r \times 2r$ matrix in the bracket, and the residue is defined as
\begin{equation}
\text{Res}_{W_n}(f (x_{i}))=\frac{1}{(2\pi i)^{u_{n}}}
\oint\frac{f (x_{i})}{\partial_{1} W
\cdots \partial_{u_{n}}W} \bigg|_{x_{j}=0\ (j>u_{n})} 
dx_{1}\dotsb dx_{u_{n}} \ .
\end{equation}
The expressions for the charges allow one to determine the intersection
index as~\cite{Walcher:2004tx}
\begin{equation}\label{index}
\text{Tr}(-1)^{F}=\left\langle \text{ch}(\widehat{Q},\hat{\gamma
}),\text{ch}(Q,\gamma )\right\rangle \ , 
\end{equation}
where 
\begin{equation}\label{pairing}
\left\langle \text{ch}(\widehat{Q},\hat{\gamma}),
\text{ch}(Q,\gamma)\right\rangle = 
\frac{1}{H} \sum_{n=0}^{H-1} \sum_{\alpha,\beta} 
\text{ch}(\widehat{Q},\hat{\gamma})(|n;\alpha\rangle)
\frac{1}{\prod_{\frac{n}{h_{i}}\notin \mathbb{Z}}(1-\omega^{w_i n})}
\eta_{n}^{\alpha \beta^{*}}
\text{ch}(Q,\gamma)(|n;\beta\rangle)^{*} \ .
\end{equation}
Here, $\eta_{n}^{\alpha \beta}$ is the inverse of the closed
topological string metric,
\begin{equation}
\eta_{\alpha \beta}^{n}=\text{Res}_{W_{n}}
(\phi_{n}^{\alpha}\phi_{n}^{\beta})
\end{equation}
and $\beta^{*}$ denotes the label of the field conjugate to
$\phi_{n}^{\beta}$ (which is given by 
$\beta^{*}_{i} = h_{i}-2-\beta_{i}$). Note that the
formula~\eqref{pairing} has been slightly modified in comparison
to~\cite{Walcher:2004tx} by introducing the conjugate label. 

For the factorisations we are considering here, namely products of
generalised permutation factorisations and tensor product
factorisations, the charge formula~\eqref{charge} factorises, so we
can focus on the case of just two factors and a generalised
permutation factorisation $Q_{\eta}$. In the untwisted sector
($n=0\mod r_{1}r_{2}d$), the charge is given by
\begin{equation}
\text{ch} (Q_{\eta},\gamma) (|n;(r_{1}m-1,r_{2} (d-m)-1)\rangle) = 
\frac{1}{d} \eta^{m}\ ,
\end{equation}
where $m=1,\dotsc ,d-1$. In the twisted sectors we only have a
non-zero contribution if $n\not= 0 \mod r_{i}d$ for $i=1,2$, and it is
given by
\begin{equation}
\text{ch} (Q_{\eta },\gamma) (|n;0\rangle) = \text{Str} (\gamma^{n})\ \ . 
\end{equation}
Using these results in~\eqref{index}, we have checked in all relevant
examples that the result agrees with the intersection
matrices obtained in section~\ref{spectra}.

\section{Application to Gepner models}

After these preparations we are now in a position to analyse the $147$
different Fermat type Calabi-Yau manifolds in detail. Before we begin
with this analysis we should explain more precisely what we expect,
and what we are looking for. 

\subsection{RR vector spaces and RR lattices}

The D-branes we are interested in carry RR charges. In general,
different D-branes carry different RR charges, and there is a whole
vector space of such charges. The dimension of this vector space
equals the dimension of the even cohomology of the corresponding
Calabi-Yau manifold. (Since we are only considering B-type D-branes
here, only the even dimensional cohomology is of relevance.) For all
the $147$ Fermat type Calabi-Yau manifolds this dimension is known
\cite{Lutken:1988hc,Lynker:1988fs,Candelas:1989hd}.\footnote{See also
the 
URL~\texttt{http://hep.itp.tuwien.ac.at/{\~{}}kreuzer/CY}.}

One natural objective we may have is to find a set of D-branes 
whose charges form a basis for this vector space. This is what we mean
by `spanning the RR vector space' in the following. From the conformal
field theory point of view, this question was analysed in 
section~2, where we saw that for $31$ models the RR vector space is 
not spanned by tensor product or permutation branes. From the point of
view of the matrix factorisations, the condition that a set of
factorisations spans the RR vector space can also be easily
formulated: it simply means that the rank of its intersection matrix 
agrees with the dimension of the even
cohomology. With the explicit formulae for the intersection matrices,
this condition can be easily tested in each case. Given that we know
which factorisations correspond to tensor product and permutation
branes, we thus expect that there are $31$ models for which these
factorisations do not span the RR vector space.  
\medskip

The RR charges of these theories do not just form a vector space, but
they actually form a lattice. This reflects the fact that RR
charges are quantised. We can therefore ask a more detailed question:
which D-branes do not just generate the RR vector space, but actually
`span the RR lattice'. From the conformal field theory point of view, 
this question is not straightforward, since we do not know how to
describe the D-branes that span the RR vector space for $31$
models. However, from the point of view of the matrix factorisations
this question can again be easily analysed: a set of factorisations
spans the RR lattice if its intersection matrix contains a submatrix
with maximal rank (equalling the dimension of the even cohomology)
that has determinant equal to $1$! Given the explicit formulae for the
various  intersection matrices we can fairly systematically look for 
factorisations which have this property.

\subsection{Explicit results}

With this in mind we have analysed the matrix factorisations for all
$147$ models. In each case we have first tried to find tensor product
and conventional permutation factorisations that span the RR vector
space. This was possible in all but $32$ cases; these $32$ cases are
described in table~2. 

\small
\renewcommand{\arraystretch}{1.1}
\begin{longtable}{lcccc}
\caption{\normalsize Models that require generalised permutation
factorisations in order to span the RR vector space.} \\
\hline
Calabi-Yau & Gepner model & RR & RR  & RR \\
 &                   &     dim     & vector space       & charge lattice \\
\hline\hline 
\endfirsthead
\caption[]{\normalsize(continued)} \\
\hline
Calabi-Yau & Gepner model & RR & RR  & RR \\
 &                   &     dim      & vector space       & charge lattice \\
\hline\hline 
\endhead
$\mathbb{P}_{\left(3,3,4,6,8\right)}[24]$ & $\left(6,6,4,2,1\right)$ &
16 & $ (\widetilde{35})_{0,2}$ &
$(12)(\widetilde{35})_{0,2}$ \\
$\mathbb{P}_{\left(1,3,6,10,10\right)}[30]$ &
$\left(28,8,3,1,1\right)$ & 40
& $(\widetilde{23})(45)_{0,2,4,6;0,2}$ & $(\widetilde{23})(45)_{0,2,4,6;0,2}$ \\
$\mathbb{P}_{\left(3,5,6,6,10\right)}[30]$ & $\left(8,4,3,3,1\right)$
& 32 & $(34)(\widetilde{25})_{0,2,4,6;0,2}$ &
$(34)(\widetilde{25})_{0,2,4,6;0,2}$ \\
$\mathbb{P}_{\left(1,1,4,12,18\right)}[36]$ &
$\left(34,34,7,1,0\right)$ & 16 & $(\widetilde{34})_{0,2}$ &
$(12)(\widetilde{34})_{0,2}$ \\
$\mathbb{P}_{\left(2,4,9,9,12\right)}[36]$ & $\left(16,7,2,2,1\right)$
& 40 & $(\widetilde{25})(34)_{0,2,4}$ &
$(\widetilde{25})(34)_{0,2,4}$ \\
$\mathbb{P}_{\left(1,3,12,16,16\right)}[48]$ &
$\left(46,14,2,1,1\right)$ & 54 & $(\widetilde{23})(45)_{0,2,4}$ &
$(\widetilde{23})(45)_{0,2,4}$ \\
$\mathbb{P}_{\left(1,2,6,18,27\right)}[54]$ &
$\left(52,25,7,1,0\right)$ & 22 & $(\widetilde{34})_{0,2}$ &
$(\widetilde{34})_{0,2}$ \\
$\mathbb{P}_{\left(1,4,5,20,30\right)}[60]$ &
$\left(58,13,10,1,0\right)$ & 24 & $(\widetilde{24})_{0,2}$ &
$(\widetilde{13})(\widetilde{24})_{0,2}$ \\
$\mathbb{P}_{\left(2,3,15,20,20\right)}[60]$ &
$\left(28,18,2,1,1\right)$ & 64 & $(\widetilde{23})(45)_{0,2,4}$ &
$(\widetilde{23})(45)_{0,2,4}$ \\
$\mathbb{P}_{\left(3,3,4,20,30\right)}[60]$ &
$\left(18,18,13,1,0\right)$ & 24 & $(\widetilde{34})_{0,2}$ &
$(12)(\widetilde{34})_{0,2}$ \\
$\mathbb{P}_{\left(3,5,12,20,20\right)}[60]$ &
$\left(18,10,3,1,1\right)$ & 64
& $(\widetilde{13})(45)_{0,2,4,6;0,2}$ & $(\widetilde{13})(45)_{0,2,4,6;0,2}$ \\
$\mathbb{P}_{\left(3,10,12,15,20\right)}[60]$ &
$\left(18,4,3,2,1\right)$ & 32 & $(\widetilde{25})_{0,2}$ &
$(\widetilde{25})_{0,2}$ \\
$\mathbb{P}_{\left(4,6,15,15,20\right)}[60]$ &
$\left(13,8,2,2,1\right)$ & 56 & $(\widetilde{15})(34)_{0,2,4}$ &
$(\widetilde{15})(34)_{0,2,4}$ \\
$\mathbb{P}_{\left(1,3,8,24,36\right)}[72]$ &
$\left(70,22,7,1,0\right)$ & 32 & $(\widetilde{34})_{0,2}$ &
$(\widetilde{12})(\widetilde{34})_{0,2}$ \\
$\mathbb{P}_{\left(1,6,21,28,28\right)}[84]$ &
$\left(82,12,2,1,1\right)$ & 82 & $(45)_{0,2}-(\widetilde{23})$ &
$(\widetilde{12})(45)_{0,2}-(\widetilde{23})(45)$ \\
$\mathbb{P}_{\left(3,4,7,28,42\right)}[84]$ &
$\left(26,19,10,1,0\right)$ & 34 & $(15)(\widetilde{24})_{0,2}$ &
$(15)(\widetilde{24})_{0,2}$ \\
$\mathbb{P}_{\left(3,4,21,28,28\right)}[84]$ &
$\left(26,19,2,1,1\right)$ & 84 & $(\widetilde{13})(45)_{0,2,4}$ &
$(\widetilde{13})(45)_{0,2,4}$ \\
$\mathbb{P}_{\left(2,3,10,30,45\right)}[90]$ &
$\left(43,28,7,1,0\right)$ & 38 & $(\widetilde{34})_{0,2}$ &
$(\widetilde{34})_{0,2}$ \\
$\mathbb{P}_{\left(1,4,20,25,50\right)}[100]$ &
$\left(98,23,3,2,0\right)$ & 68
& $(\widetilde{23})(45)_{0,2,4,6;0}$ & $(\widetilde{23})(45)_{0,2,4,6;0}$ \\
$\mathbb{P}_{\left(1,6,14,42,63\right)}[126]$ &
$\left(124,19,7,1,0\right)$ & 58 &
$(\widetilde{34})_{0,2}-(\widetilde{24})_{0,2}$
& $(\widetilde{34})_{0,2}-(\widetilde{24})_{0,2}$ \\
$\mathbb{P}_{\left(1,9,20,60,90\right)}[180]$ &
$\left(178,18,7,1,0\right)$ & 86 & $(25)(\widetilde{34})_{0,2}$ &
$(25)(\widetilde{34})_{0,2}$ \\
$\mathbb{P}_{\left(4,5,36,45,90\right)}[180]$ 
& $\left(43,34,3,2,0\right)$ & 112 
& $(\widetilde{13})(45)_{0,2,4,6;0}$ & $(\widetilde{13})(45)_{0,2,4,6;0}$ \\
$\mathbb{P}_{\left(2,9,22,66,99\right)}[198]$ &
$\left(97,20,7,1,0\right)$ & 92 & $(25)(\widetilde{34})_{0,2}$ &
$(25)(\widetilde{34})_{0,2}$ \\
$\mathbb{P}_{\left(6,14,15,70,105\right)}[210]$ 
& $\left(33,13,12,1,0\right)$ & 88 
& $(\widetilde{24})(35)_{0,2}$ & $(\widetilde{24})(35)_{0,2}$ \\
$\mathbb{P}_{\left(1,8,27,72,108\right)}[216]$ 
& $\left(214,25,6,1,0\right)$ & 98 
& $(\widetilde{24})(35)_{0,2}$ & $(\widetilde{24})(35)_{0,2}$ \\
$\mathbb{P}_{\left(3,8,33,88,132\right)}[264]$ 
& $\left(86,31,6,1,0\right)$ & 116 
& $(\widetilde{24})(35)_{0,2}$ & $(\widetilde{24})(35)_{0,2}$ \\
$\mathbb{P}_{\left(1,6,42,98,147\right)}[294]$ 
& $\left(292,47,5,1,0\right)$ & 96 
& $(\widetilde{23})_{0,\ldots,10}$ & $(\widetilde{23})_{0,\ldots,10}$ \\
$\mathbb{P}_{\left(3,22,30,110,165\right)}[330]$ 
& $\left(108,13,9,1,0\right)$ & 120 
& $(\widetilde{24})_{0,2}$ & $(\widetilde{24})_{0,2}$ \\
$\mathbb{P}_{\left(1,18,38,114,171\right)}[342]$ 
& $\left(340,17,7,1,0\right)$ & 144 
& $(\widetilde{34})_{0,2}$ & $(\widetilde{34})_{0,2}$ \\
$\mathbb{P}_{\left(6,7,78,182,273\right)}[546]$ 
& $\left(89,76,5,1,0\right)$ & 168 
& $(\widetilde{13})_{0,\ldots,10}$ & $(\widetilde{13})_{0,\ldots,10}$ \\
$\mathbb{P}_{\left(1,14,90,210,315\right)}[630]$ 
& $\left(628,43,5,1,0\right)$ & 192 
& $(\widetilde{24})_{0,2}$ & $(\widetilde{24})_{0,2}$ \\
$\mathbb{P}_{\left(3,14,102,238,357\right)}[714]$ 
& $\left(236,49,5,1,0\right)$ & 216 
& $(\widetilde{24})_{0,2}$ & $(\widetilde{24})_{0,2}$ \\[0.1cm]
\hline\hline \\
\end{longtable}
\renewcommand{\arraystretch}{1}
\normalsize
Most of the conventions used in this table should be clear; a
more detailed explanation of the conventions is given in the
appendix. 

This list of models agrees with that of table~1, except for the model 
$\mathbb{P}_{\left(1,6,21,28,28\right)}[84]$, for which the matrix
factorisation analysis predicts that a generalised permutation
factorisation is required in the factor $(\widetilde{23})$, while the
conformal field theory analysis predicts that all RR charges can be
accounted for in terms of the usual tensor product and permutation branes. 
This mismatch is easily resolved: the generalised permutation
factorisation $(\widetilde{23})$ is of the form
\begin{equation}\label{resten1}
Q \ : \qquad J = x_2^{h_{2}/2} + i x_3^{h_{3}/2}\ , 
\qquad E = x_2^{h_{2}/2} - i x_3^{h_{3}/2}  \ ,
\end{equation}
or 
\begin{equation}\label{resten2}
Q^r \ : \qquad J = x_2^{h_{2}/2} - i x_3^{h_{3}/2}\ , 
\qquad E = x_2^{h_{2}/2} + i x_3^{h_{3}/2}  \ .
\end{equation}
These two factorisations describe anti-branes of one another, and one
can easily see that their direct sum $Q\oplus Q^r$ is equivalent (in
the sense of \cite{Herbst:2004jp}) to the tensor product factorisation
with $J_2=E_2=x_2^{h_{2}/2}$, $J_3=E_3=x_3^{h_{3}/2}$. Thus these two
factorisations describe in fact the two resolved (short orbit)
tensor product D-branes! 

This identification of certain rank~$1$ factorisations with resolved
tensor product D-branes actually occurs more generally,
whenever $d$ (and thus $h_2$ and $h_3$) is even. Then one always has
the two rank $1$ factorisations (\ref{resten1}) and 
(\ref{resten2}) as above, and by the same argument they always
correspond to resolved tensor product D-branes. In general, these
factorisations are just two special cases of a whole set of
(generalised) permutation factorisations; in fact they correspond to
the factorisations where $J$ contains precisely every second 
$d^{\text{th}}$ root of $-1$ (while $E$ is the product of the other
factors). What is special about the case when $d=2$ is the only common
divisor, is that {\it all} possible generalised permutation
factorisations are of this type. 

The simplest case for which all the `generalised permutation
factorisations' correspond to resolved tensor product branes therefore occurs
when one of the two factors has $h=2$ (and thus $k=0$). This is in
fact  also clear from the conformal field theory perspective since the level 
$k=0$ factor is trivial, and thus cannot be involved in any `new'
construction. Since these factorisations appear quite frequently
and are obviously not new, we have treated them as conventional
tensor product constructions. Apart from these cases, the model
$\mathbb{P}_{\left(1,6,21,28,28\right)}[84]$ is the only example where
a generalised permutation factorisation with $d=2$ occurs. In
particular, it is then clear that for the remaining $31$ theories
the generalised permutation factorisations cannot be interpreted as 
resolved tensor product branes. Thus the matrix factorisation
analysis agrees beautifully with the conformal field theory analysis
of section~2. However, now we can actually specify the D-branes that
are required to account for the charges: they are described by
generalised permutation factorisations. 
Given the similarity to the usual permutation factorisations, it seems
appropriate to call the corresponding branes `generalised permutation
branes'.  
\medskip

From the point of view of the matrix factorisation description we
can also analyse which factorisations are required in order to span
the RR charge lattice (not just the RR vector space). We have analysed
this question for all the $147$ models in detail, and we have found
that the only factorisations that are required in addition to the
tensor product and permutation factorisations are again the
generalised permutation factorisations described above. The actual
constructions that do the job in all cases are spelled out in table~2
and table~3 (in the appendix).

\section{Summary}

In summary, we have therefore shown that in general the tensor product
and permutation branes do not account for all RR charges of Gepner
models. This result could be obtained using either a direct conformal
field theory analysis, or the results from the matrix factorisations
description of D-branes. Using the latter approach, we could
furthermore identify, at least for the $147$ Fermat type Calabi-Yau
manifolds, what additional constructions are needed: the only
additional D-branes that are required are generalised permutation
branes that should exist whenever the relevant (shifted) levels have a
non-trivial common factor. We were also able to show that the
corresponding generalised permutation factorisations generate the full
RR charge lattice.

The generalised permutation branes that appear in our
analysis are very reminiscent of the generalised permutation branes
for products of SU(2) WZW-models for which evidence was recently found
in \cite{Fredenhagen:2005an}. The matrix factorisation description also
determines various properties of these D-branes, such as their
charges, the topological open string spectrum, {\it etc.} This should
help to construct these D-branes in conformal field theory.

\section*{Acknowledgements}

This research has been partially supported by 
the Swiss National Science Foundation and the Marie Curie network
`Constituents, Fundamental Forces and Symmetries of the Universe'
(MRTN-CT-2004-005104). The work of S.F.\ was supported by the Max
Planck Society and the Max Planck Institute for Gravitational Physics
in Golm. We thank Ilka Brunner for useful discussions. This paper
is largely based on the Diploma thesis of one of us (C.C.)
\cite{Caviezel}.

\begin{appendix}

\section{The generating matrix factorisations}

In this appendix we describe the matrix factorisations that span the RR
vector space and the RR charge lattice for all $147-32$ A-type Gepner
models. (The remaining $32$ cases were already described in table~2.)

In the following table, the first three entries should be
self-explanatory. In the fourth and fifth column we have described the
matrix factorisations that generate the RR vector space, and the RR
charge lattice, respectively. A few words containing conventions are
in order. 
\begin{list}{(\roman{enumi})}{\usecounter{enumi}}
\item RS stands for Recknagel-Schomerus construction,
{\it i.e.} for the simplest tensor product factorisations whose
intersection form was given at the end of section~3.1. 
\item Similarly, ${(12)}$ stands for the (12)-permutation
factorisation (with tensor product factorisations for the factors
$3,4$ and $5$). In the same vain, ${(12)(34)}$ denotes the  
permutation factorisation, where we have a transposition in the
factors (12) and the factors (34), {\it etc}. 
\item By $(45)_{0,2}$ we mean that one has to consider the
(45)-permutation factorisations with two different values for $\eta$: 
$\eta=e^{\frac{-\pi i (M+1)}{d}}$ with $M=0,2$. \\
Similarly,
by $(23)(45)_{0,\ldots,10;0,2}$ we mean that one takes the 
(23)-permutation factorisation with $M_{23}=0,2,\ldots,10$, and
the (45)-permutation factorisation with $M_{45}=0,2$. (For ease of
notation we write $(23)(45)_{0,2,4}$ for $(23)(45)_{0,2,4;0,2,4}$.)  
\item By $(45)$$-$$(35)$ we mean the set of factorisations which have a
(45)-permutation factorisation (but are of tensor type in the first,
second and third factor), together with the set of factorisations that
are (35)-permutation but tensor in the first, second and fourth
factor. 
\end{list}

\noindent We have furthermore denoted generalised permutation
factorisations by a tilde; for example $(\widetilde{12})$ denotes the
generalised permutation factorisation in the first two factors. There
is one exception to this rule: as was explained at the end of
section~4.2, the `generalised' permutation factorisations that involve
a trivial factor with $k=0$ (usually the fifth factor), do not
describe novel D-branes, but rather correspond to resolved 
tensor product branes. Thus for example in the model
$\mathbb{P}_{\left(1,1,1,3,6\right)}[12]$, whose corresponding levels
are $(k_1,k_2,k_3,k_4,k_5)=(10,10,10,2,0)$ the generalised permutation
factorisation involving the last two factors describes in fact a usual
tensor product brane. 

The models included below have the property that the RR vector space
is spanned by tensor product or permutation branes. However, even for these
models it is clear that some require generalised permutation branes to 
account for the full RR charge lattice! 

\small
\renewcommand{\arraystretch}{1.1}
\begin{longtable}{lcccc}
\caption{\normalsize The models where the RR vector space is generated by 
tensor product and conventional permutation branes.} \\
\hline
Calabi-Yau & Gepner model & RR & RR  & RR \\
  &                   &    dim      & vector space       & charge lattice \\
\hline\hline 
\endfirsthead
\caption[]{(\normalsize continued)} \\
\hline
Calabi-Yau & Gepner model & RR & RR  & RR \\
  &                   &    dim      & vector space       & charge lattice \\
\hline\hline 
\endhead
$\mathbb{P}_{\left(1,1,1,1,1\right)}[5]$ & $\left(3,3,3,3,3\right)$ &
4 & RS &
${(12)(34)}$ \\
$\mathbb{P}_{\left(1,1,1,1,2\right)}[6]$ & $\left(4,4,4,4,1\right)$ &
4 & {RS} &
${(12)(34)}$ \\
$\mathbb{P}_{\left(1,1,1,1,4\right)}[8]$ & $\left(6,6,6,6,0\right)$ &
4 & {RS} &
${(12)(34)}$ \\
$\mathbb{P}_{\left(1,1,2,2,2\right)}[8]$ & $\left(6,6,2,2,2\right)$ &
6 & {RS} &
${(12)(34)}$ \\
$\mathbb{P}_{\left(1,1,1,3,3\right)}[9]$ & $\left(7,7,7,1,1\right)$ &
10 & $(45)_{0,2}$ &
$(12)(45)_{0,2}$ \\
$\mathbb{P}_{\left(1,1,1,2,5\right)}[10]$ & $\left(8,8,8,3,0\right)$ &
4 & {RS} & {RS} \\
$\mathbb{P}_{\left(1,1,1,3,6\right)}[12]$ &
$\left(10,10,10,2,0\right)$ & 8 & ${(45)}$ &
${(12)(45)}$ \\
$\mathbb{P}_{\left(1,1,2,2,6\right)}[12]$ & $\left(10,10,4,4,0\right)$
& 6 & {RS} &
${(12)(34)}$ \\
$\mathbb{P}_{\left(1,1,2,4,4\right)}[12]$ & $\left(10,10,4,1,1\right)$
& 12 & $(45)_{0,2}$ &
$(12)(45)_{0,2}$ \\
$\mathbb{P}_{\left(1,1,3,3,4\right)}[12]$ & $\left(10,10,2,2,1\right)$
& 12 & $(34)_{0,2,4}$ &
$(12)(34)_{0,2,4}$ \\
$\mathbb{P}_{\left(1,2,2,3,4\right)}[12]$ & $\left(10,4,4,2,1\right)$
& 6 & {RS} &
${(23)}$ \\
$\mathbb{P}_{\left(1,2,3,3,3\right)}[12]$ & $\left(10,4,2,2,2\right)$
& 8 & {RS} &
${(\widetilde{12})(34)}$ \\
$\mathbb{P}_{\left(2,2,2,3,3\right)}[12]$ & $\left(4,4,4,2,2\right)$ &
14 & $(45)_{0,2,4}$ &
$(12)(45)_{0,2,4}$ \\
$\mathbb{P}_{\left(1,2,2,2,7\right)}[14]$ & $\left(12,5,5,5,0\right)$
& 6 & {RS} &
${(23)}$ \\
$\mathbb{P}_{\left(1,1,3,5,5\right)}[15]$ & $\left(13,13,3,1,1\right)$
& 16 & $(45)_{0,2}$ &
$(12)(45)_{0,2}$ \\
$\mathbb{P}_{\left(1,3,3,3,5\right)}[15]$ & $\left(13,3,3,3,1\right)$
& 8 & {RS} &
${(23)}$ \\
$\mathbb{P}_{\left(1,1,2,4,8\right)}[16]$ & $\left(14,14,6,2,0\right)$
& 10 & ${(45)}$ &
${(12)(45)}$ \\
$\mathbb{P}_{\left(1,1,1,6,9\right)}[18]$ &
$\left(16,16,16,1,0\right)$ & 6 & {RS} &
{RS} \\
$\mathbb{P}_{\left(1,2,3,3,9\right)}[18]$ & $\left(16,7,4,4,0\right)$
& 8 & {RS} &
${(34)}$ \\
$\mathbb{P}_{\left(1,2,3,6,6\right)}[18]$ & $\left(16,7,4,1,1\right)$
& 16 & $(45)_{0,2}$ &
$(\widetilde{12})(45)_{0,2}$ \\
$\mathbb{P}_{\left(2,2,2,3,9\right)}[18]$ & $\left(7,7,7,4,0\right)$ &
10 & ${(45)}$ &
${(12)(45)}$ \\
$\mathbb{P}_{\left(1,1,4,4,10\right)}[20]$ &
$\left(18,18,3,3,0\right)$ & 16 & $(34)_{0,2,4,6}$ &
$(12)(34)_{0,2,4,6}$ \\
$\mathbb{P}_{\left(1,2,2,5,10\right)}[20]$ & $\left(18,8,8,2,0\right)$
& 14 & ${(45)}$ &
${(23)(45)}$ \\
$\mathbb{P}_{\left(1,4,5,5,5\right)}[20]$ & $\left(18,3,2,2,2\right)$
& 12 & {RS} &
${(34)}$ \\
$\mathbb{P}_{\left(2,4,4,5,5\right)}[20]$ & $\left(8,3,3,2,2\right)$ &
32 & $(23)(45)_{0,2,4,6}$ &
$(23)(45)_{0,2,4,6}$ \\
$\mathbb{P}_{\left(1,3,3,7,7\right)}[21]$ & $\left(19,5,5,1,1\right)$
& 36 & $(23)(45)_{0,\ldots,10;0,2}$ &
$(23)(45)_{0,\ldots,10;0,2}$ \\
$\mathbb{P}_{\left(1,1,2,8,12\right)}[24]$ &
$\left(22,22,10,1,0\right)$ & 8 & {RS} &
{RS} \\
$\mathbb{P}_{\left(1,1,4,6,12\right)}[24]$ &
$\left(22,22,4,2,0\right)$ & 18 & ${(45)}$$-$${(35)}$ &
${(12)(45)}$$-$${(12)(35)}$ \\
$\mathbb{P}_{\left(1,1,6,8,8\right)}[24]$ & $\left(22,22,2,1,1\right)$
& 24 & $(45)_{0,2}$ &
$(12)(45)_{0,2}$ \\
$\mathbb{P}_{\left(1,2,3,6,12\right)}[24]$ &
$\left(22,10,6,2,0\right)$ & 14 & ${(45)}$ &
${(\widetilde{12})(45)}$ \\
$\mathbb{P}_{\left(1,3,4,4,12\right)}[24]$ & $\left(22,6,4,4,0\right)$
& 14 & ${(25)}$ &
${(25)(34)}$ \\
$\mathbb{P}_{\left(1,3,4,8,8\right)}[24]$ & $\left(22,6,4,1,1\right)$
& 22 & $(45)_{0,2}$ &
$(\widetilde{12})(45)_{0,2}$ \\
$\mathbb{P}_{\left(1,3,6,6,8\right)}[24]$ & $\left(22,6,2,2,1\right)$
& 18 & $(34)_{0,2,4}$ &
$(34)_{0,2,4}$ \\
$\mathbb{P}_{\left(2,3,3,4,12\right)}[24]$ & $\left(10,6,6,4,0\right)$
& 14 & ${(45)}$ &
${(23)(45)}$ \\
$\mathbb{P}_{\left(2,3,3,8,8\right)}[24]$ & $\left(10,6,6,1,1\right)$
& 40 & $(23)(45)_{0,\ldots,12;0,2}$ &
$(23)(45)_{0,\ldots,12;0,2}$ \\
$\mathbb{P}_{\left(1,2,4,7,14\right)}[28]$ &
$\left(26,12,5,2,0\right)$ & 18 & ${(45)}$ &
${(\widetilde{12})(45)}$ \\
$\mathbb{P}_{\left(1,1,3,10,15\right)}[30]$ &
$\left(28,28,8,1,0\right)$ & 12 & ${(35)}$ &
${(12)(35)}$ \\
$\mathbb{P}_{\left(1,2,2,10,15\right)}[30]$ &
$\left(28,13,13,1,0\right)$ & 10 & {RS} &
${(23)}$ \\
$\mathbb{P}_{\left(1,2,6,6,15\right)}[30]$ &
$\left(28,13,3,3,0\right)$ & 20 & $(34)_{0,2,4,6}$ &
$(34)_{0,2,4,6}$ \\
$\mathbb{P}_{\left(1,3,5,6,15\right)}[30]$ & $\left(28,8,4,3,0\right)$
& 16 & ${(35)}$ &
${(\widetilde{12})(35)}$ \\
$\mathbb{P}_{\left(2,2,5,6,15\right)}[30]$ &
$\left(13,13,4,3,0\right)$ & 16 & ${(35)}$ &
${(12)(35)}$ \\
$\mathbb{P}_{\left(2,3,5,5,15\right)}[30]$ & $\left(13,8,4,4,0\right)$
& 16 & ${(25)}$ &
${(34)(25)}$ \\
$\mathbb{P}_{\left(2,3,5,10,10\right)}[30]$ &
$\left(13,8,4,1,1\right)$ & 26 & $(45)_{0,2}$ &
$(45)_{0,2}$ \\
$\mathbb{P}_{\left(1,2,3,12,18\right)}[36]$ &
$\left(34,16,10,1,0\right)$ & 12 & {RS} &
{RS} \\
$\mathbb{P}_{\left(1,2,6,9,18\right)}[36]$ &
$\left(34,16,4,2,0\right)$ & 26 & ${(45)}$$-$${(35)}$ &
${(\widetilde{12})(45)}$$-$${(\widetilde{12})(35)}$ \\
$\mathbb{P}_{\left(1,2,9,12,12\right)}[36]$ &
$\left(34,16,2,1,1\right)$ & 34 & $(45)_{0,2}$ &
$(\widetilde{12})(45)_{0,2}$ \\
$\mathbb{P}_{\left(1,4,4,9,18\right)}[36]$ & $\left(34,7,7,2,0\right)$
& 40 & $(23)(45)_{0,\ldots,14;0}$ &
$(23)(45)_{0,\ldots,14;0}$ \\
$\mathbb{P}_{\left(2,3,4,9,18\right)}[36]$ &
$\left(16,10,7,2,0\right)$ & 22 & ${(45)}$ &
${(\widetilde{13})(45)}$ \\
$\mathbb{P}_{\left(1,1,8,10,20\right)}[40]$ &
$\left(38,38,3,2,0\right)$ & 24 & ${(45)}$ &
${(12)(45)}$ \\
$\mathbb{P}_{\left(1,4,5,10,20\right)}[40]$ &
$\left(38,8,6,2,0\right)$ & 26 & ${(45)}$$-$${(25)}$ &
${(\widetilde{13})(45)}$$-$${(\widetilde{13})(25)}$ \\
$\mathbb{P}_{\left(2,5,5,8,20\right)}[40]$ & $\left(18,6,6,3,0\right)$
& 16 & {RS} &
${(23)}$ \\
$\mathbb{P}_{\left(1,3,3,14,21\right)}[42]$ &
$\left(40,12,12,1,0\right)$ & 14 & {RS} &
${(23)}$ \\
$\mathbb{P}_{\left(1,6,7,7,21\right)}[42]$ & $\left(40,5,4,4,0\right)$
& 18 & {RS} &
${(34)}$ \\
$\mathbb{P}_{\left(1,6,7,14,14\right)}[42]$ &
$\left(40,5,4,1,1\right)$ & 36 & $(45)_{0,2}$ &
$(45)_{0,2}$ \\
$\mathbb{P}_{\left(2,2,3,14,21\right)}[42]$ &
$\left(19,19,12,1,0\right)$ & 16 & ${(35)}$ &
${(12)(35)}$ \\
$\mathbb{P}_{\left(2,6,6,7,21\right)}[42]$ & $\left(19,5,5,4,0\right)$
& 36 & $(23)(45)_{0,\ldots,10;0}$ &
$(23)(45)_{0,\ldots,10;0}$ \\
$\mathbb{P}_{\left(1,5,9,15,15\right)}[45]$ &
$\left(43,7,3,1,1\right)$ & 40 & $(45)_{0,2}$ &
$(45)_{0,2}$ \\
$\mathbb{P}_{\left(1,1,6,16,24\right)}[48]$ &
$\left(46,46,6,1,0\right)$ & 20 & ${(35)}$ &
${(12)(35)}$ \\
$\mathbb{P}_{\left(1,3,4,16,24\right)}[48]$ &
$\left(46,14,10,1,0\right)$ & 18 & ${(25)}$ &
${(\widetilde{13})(25)}$ \\
$\mathbb{P}_{\left(1,3,8,12,24\right)}[48]$ &
$\left(46,14,4,2,0\right)$ & 34 & ${(45)}$$-$${(35)}$ &
${(\widetilde{12})(45)}$$-$${(\widetilde{12})(35)}$ \\
$\mathbb{P}_{\left(2,3,3,16,24\right)}[48]$ &
$\left(22,14,14,1,0\right)$ & 16 & {RS} &
${(23)}$ \\
$\mathbb{P}_{\left(1,2,12,15,30\right)}[60]$ &
$\left(58,28,3,2,0\right)$ & 36 & ${(45)}$ &
${(\widetilde{12})(45)}$ \\
$\mathbb{P}_{\left(1,3,6,20,30\right)}[60]$ &
$\left(58,18,8,1,0\right)$ & 22 & ${(35)}$ &
${(35)}$ \\
$\mathbb{P}_{\left(1,4,10,15,30\right)}[60]$ &
$\left(58,13,4,2,0\right)$ & 42 & ${(45)}$$-$${(35)}$ &
${(\widetilde{12})(45)}$$-$${(\widetilde{12})(35)}$ \\
$\mathbb{P}_{\left(1,4,15,20,20\right)}[60]$ &
$\left(58,13,2,1,1\right)$ & 54 & $(45)_{0,2}$ &
$(45)_{0,2}$ \\
$\mathbb{P}_{\left(1,5,12,12,30\right)}[60]$ &
$\left(58,10,3,3,0\right)$ & 52 & $(34)(25)_{0,2,4,6;0}$ &
$(34)(25)_{0,2,4,6;0}$ \\
$\mathbb{P}_{\left(1,12,12,15,20\right)}[60]$ &
$\left(58,3,3,2,1\right)$ & 48 & $(23)_{0,2,4,6}$ &
$(23)_{0,2,4,6}$ \\
$\mathbb{P}_{\left(2,3,5,20,30\right)}[60]$ &
$\left(28,18,10,1,0\right)$ & 22 & ${(25)}$ &
${(\widetilde{13})(25)}$ \\
$\mathbb{P}_{\left(2,3,10,15,30\right)}[60]$ &
$\left(28,18,4,2,0\right)$ & 42 & ${(45)}$$-$${(35)}$ &
${(\widetilde{13})(45)}$$-$${(35)}$ \\
$\mathbb{P}_{\left(3,5,10,12,30\right)}[60]$ &
$\left(18,10,4,3,0\right)$ & 28 & ${(35)}$ &
${(35)}$ \\
$\mathbb{P}_{\left(4,5,6,15,30\right)}[60]$ &
$\left(13,10,8,2,0\right)$ & 38 & ${(45)}$$-$${(35)}$ &
${(45)}$$-$${(35)}$ \\
$\mathbb{P}_{\left(2,3,6,22,33\right)}[66]$ &
$\left(31,20,9,1,0\right)$ & 20 & {RS} &
{RS} \\
$\mathbb{P}_{\left(1,10,10,14,35\right)}[70]$ &
$\left(68,5,5,3,0\right)$ & 48 & $(23)_{0,\ldots,10}$ &
$(23)_{0,\ldots,10}$ \\
$\mathbb{P}_{\left(2,5,14,14,35\right)}[70]$ &
$\left(33,12,3,3,0\right)$ & 56 & $(34)(25)_{0,2,4,6;0}$ &
$(34)(25)_{0,2,4,6;0}$ \\
$\mathbb{P}_{\left(1,2,9,24,36\right)}[72]$ &
$\left(70,34,6,1,0\right)$ & 30 & ${(35)}$ &
${(\widetilde{12})(35)}$ \\
$\mathbb{P}_{\left(1,8,9,18,36\right)}[72]$ &
$\left(70,7,6,2,0\right)$ & 40 & ${(45)}$ &
${(45)}$ \\
$\mathbb{P}_{\left(1,6,6,26,39\right)}[78]$ &
$\left(76,11,11,1,0\right)$ & 48 & $(23)_{0,\ldots,22}$ &
$(23)_{0,\ldots,22}$ \\
$\mathbb{P}_{\left(1,1,12,28,42\right)}[84]$ &
$\left(82,82,5,1,0\right)$ & 24 & {RS} &
{RS} \\
$\mathbb{P}_{\left(1,6,7,28,42\right)}[84]$ &
$\left(82,12,10,1,0\right)$ & 32 & ${(25)}$ &
${(\widetilde{13})(25)}$ \\
$\mathbb{P}_{\left(1,6,14,21,42\right)}[84]$ &
$\left(82,12,4,2,0\right)$ & 62 & ${(45)}$$-$${(35)}$$-$${(25)}$ &
${(\widetilde{12})(45)}$$-$${(\widetilde{12})(35)}$$-$${(25)}$ \\
$\mathbb{P}_{\left(2,7,12,21,42\right)}[84]$ &
$\left(40,10,5,2,0\right)$ & 48 & ${(45)}$ &
${(45)}$ \\
$\mathbb{P}_{\left(2,12,21,21,28\right)}[84]$ &
$\left(40,5,2,2,1\right)$ & 72 & $(34)_{0,2,4}$ &
$(34)_{0,2,4}$ \\
$\mathbb{P}_{\left(3,4,14,21,42\right)}[84]$ &
$\left(26,19,4,2,0\right)$ & 58 & ${(45)}$$-$${(35)}$ &
${(45)}$$-$${(35)}$ \\
$\mathbb{P}_{\left(1,5,9,30,45\right)}[90]$ &
$\left(88,16,8,1,0\right)$ & 36 & ${(35)}$ &
${(\widetilde{12})(35)}$ \\
$\mathbb{P}_{\left(2,10,15,18,45\right)}[90]$ &
$\left(43,7,4,3,0\right)$ & 40 & ${(35)}$ &
${(35)}$ \\
$\mathbb{P}_{\left(1,3,12,32,48\right)}[96]$ &
$\left(94,30,6,1,0\right)$ & 38 & ${(35)}$ &
${(35)}$ \\
$\mathbb{P}_{\left(1,10,22,22,55\right)}[110]$ &
$\left(108,9,3,3,0\right)$ & 80 & $(34)_{0,2,4,6}$ &
$(34)_{0,2,4,6}$ \\
$\mathbb{P}_{\left(1,4,15,40,60\right)}[120]$ &
$\left(118,28,6,1,0\right)$ & 50 & ${(35)}$ &
${(\widetilde{12})(35)}$ \\
$\mathbb{P}_{\left(1,5,24,30,60\right)}[120]$ &
$\left(118,22,3,2,0\right)$ & 68 & ${(45)}$ &
${(45)}$ \\
$\mathbb{P}_{\left(1,15,20,24,60\right)}[120]$ &
$\left(118,6,4,3,0\right)$ & 68 & ${(35)}$$-$${(25)}$ &
${(35)}$$-$${(25)}$ \\
$\mathbb{P}_{\left(1,15,24,40,40\right)}[120]$ &
$\left(118,6,3,1,1\right)$ & 112 & $(45)_{0,2}$ &
$(45)_{0,2}$ \\
$\mathbb{P}_{\left(2,3,15,40,60\right)}[120]$ &
$\left(58,38,6,1,0\right)$ & 48 & ${(35)}$ &
${(35)}$ \\
$\mathbb{P}_{\left(3,5,12,40,60\right)}[120]$ &
$\left(38,22,8,1,0\right)$ & 46 & ${(35)}$ &
${(35)}$ \\
$\mathbb{P}_{\left(1,2,18,42,63\right)}[126]$ &
$\left(124,61,5,1,0\right)$ & 36 & {RS} &
{RS} \\
$\mathbb{P}_{\left(1,14,20,35,70\right)}[140]$ &
$\left(138,8,5,2,0\right)$ & 90 & ${(45)}$$-$${(25)}$ &
${(45)}$$-$${(25)}$ \\
$\mathbb{P}_{\left(2,5,28,35,70\right)}[140]$ &
$\left(68,26,3,2,0\right)$ & 80 & ${(45)}$ &
${(45)}$ \\
$\mathbb{P}_{\left(1,12,13,52,78\right)}[156]$ &
$\left(154,11,10,1,0\right)$ & 48 & {RS} &
{RS} \\
$\mathbb{P}_{\left(1,12,26,39,78\right)}[156]$ &
$\left(154,11,4,2,0\right)$ & 108 & ${(45)}$$-$${(35)}$ &
${(45)}$$-$${(35)}$ \\
$\mathbb{P}_{\left(1,12,39,52,52\right)}[156]$ &
$\left(154,11,2,1,1\right)$ & 144 & $(45)_{0,2}$ &
$(45)_{0,2}$ \\
$\mathbb{P}_{\left(1,3,24,56,84\right)}[168]$ &
$\left(166,54,5,1,0\right)$ & 48 & {RS} &
{RS} \\
$\mathbb{P}_{\left(1,6,21,56,84\right)}[168]$ &
$\left(166,26,6,1,0\right)$ & 70 & ${(35)}$$-$${(25)}$ &
${(35)}$$-$${(25)}$ \\
$\mathbb{P}_{\left(3,4,21,56,84\right)}[168]$ &
$\left(54,40,6,1,0\right)$ & 68 & ${(35)}$ &
${(35)}$ \\
$\mathbb{P}_{\left(2,3,30,70,105\right)}[210]$ &
$\left(103,68,5,1,0\right)$ & 60 & {RS} &
{RS} \\
$\mathbb{P}_{\left(1,10,44,55,110\right)}[220]$ &
$\left(218,20,3,2,0\right)$ & 132 & ${(45)}$$-$${(25)}$ &
${(45)}$$-$${(25)}$ \\
$\mathbb{P}_{\left(1,15,24,80,120\right)}[240]$ &
$\left(238,14,8,1,0\right)$ & 102 & ${(35)}$$-$${(25)}$ &
${(35)}$$-$${(25)}$ \\
$\mathbb{P}_{\left(1,12,39,104,156\right)}[312]$ &
$\left(310,24,6,1,0\right)$ & 134 & ${(35)}$$-$${(25)}$ &
${(35)}$$-$${(25)}$ \\
$\mathbb{P}_{\left(1,7,48,112,168\right)}[336]$ &
$\left(334,46,5,1,0\right)$ & 96 & {RS} &
{RS} \\
$\mathbb{P}_{\left(2,7,54,126,189\right)}[378]$ &
$\left(187,52,5,1,0\right)$ & 108 & {RS} &
{RS} \\
$\mathbb{P}_{\left(1,20,84,105,210\right)}[420]$ &
$\left(418,19,3,2,0\right)$ & 240 & ${(45)}$ &
${(45)}$ \\
$\mathbb{P}_{\left(3,7,60,140,210\right)}[420]$ &
$\left(138,58,5,1,0\right)$ & 120 & {RS} &
{RS} \\
$\mathbb{P}_{\left(2,33,42,154,231\right)}[462]$ &
$\left(229,12,9,1,0\right)$ & 160 & ${(25)}$ &
${(25)}$ \\
$\mathbb{P}_{\left(1,24,75,200,300\right)}[600]$ &
$\left(598,23,6,1,0\right)$ & 240 & ${(35)}$ &
${(35)}$ \\
$\mathbb{P}_{\left(1,21,132,308,462\right)}[924]$ &
$\left(922,42,5,1,0\right)$ & 276 & ${(25)}$ &
${(25)}$ \\
$\mathbb{P}_{\left(2,21,138,322,483\right)}[966]$ &
$\left(481,44,5,1,0\right)$ & 288 & ${(25)}$ &
${(25)}$ \\
$\mathbb{P}_{\left(1,42,258,602,903\right)}[1806]$ &
$\left(1804,41,5,1,0\right)$ & 504 & {RS} &
{RS} \\[0.1cm]
\hline\hline \\
\end{longtable}
\renewcommand{\arraystretch}{1}
\normalsize

\end{appendix}

\providecommand{\href}[2]{#2}\begingroup\raggedright
\endgroup

\end{document}